\documentclass[conference]{IEEEtran}
\IEEEoverridecommandlockouts
\usepackage{cite}
\usepackage{amsmath,amssymb,amsfonts}
\usepackage{bbm}
\usepackage{algorithm}
\usepackage{graphicx}
\usepackage{textcomp}
\usepackage[noend]{algpseudocode}
\usepackage{xcolor}
\def\BibTeX{{\rm B\kern-.05em{\sc i\kern-.025em b}\kern-.08em
    T\kern-.1667em\lower.7ex\hbox{E}\kern-.125emX}}

\usepackage[all]{background}
\usepackage{stackengine}
\setstackEOL{\\}
\setstackgap{L}{\normalbaselineskip}
\SetBgContents{\color{blue}{\tiny \Longstack{PREPRINT - accepted at Design Automation Conference (DAC), 2023.}}} 
\SetBgPosition{4.5cm,1cm} 
\SetBgOpacity{1.0} 
\SetBgAngle{0} 
\SetBgScale{1.8} 

\begin{document}

\title{Similarity-Based Logic Locking Against Machine Learning Attacks}

\author{\IEEEauthorblockN{Subhajit Dutta Chowdhury, Kaixin Yang, Pierluigi Nuzzo}
\IEEEauthorblockA{Ming Hsieh Department of Electrical and Computer Engineering, University of Southern California, Los Angeles, CA \\
\{duttacho, kaixinya, nuzzo\}@usc.edu}
}

\maketitle

\begin{abstract}
Logic locking is a promising technique for protecting integrated circuit designs while outsourcing their fabrication. Recently, graph neural network (GNN)-based link prediction attacks have been developed which can successfully break all the multiplexer-based locking techniques that were expected to be learning-resilient. We present SimLL, a novel similarity-based locking technique which locks a design using multiplexers and shows robustness against the existing structure-exploiting oracle-less learning-based attacks. Aiming to confuse the machine learning (ML) models, SimLL introduces key-controlled multiplexers between logic gates or wires that exhibit high levels of topological and functional similarity. Empirical results show that SimLL can degrade the accuracy of existing ML-based attacks to approximately 50\%, resulting in a negligible advantage over random guessing.
\end{abstract}


\section{Introduction}

\begin{figure}[t]
  \centering
  \includegraphics[width=0.7\columnwidth]{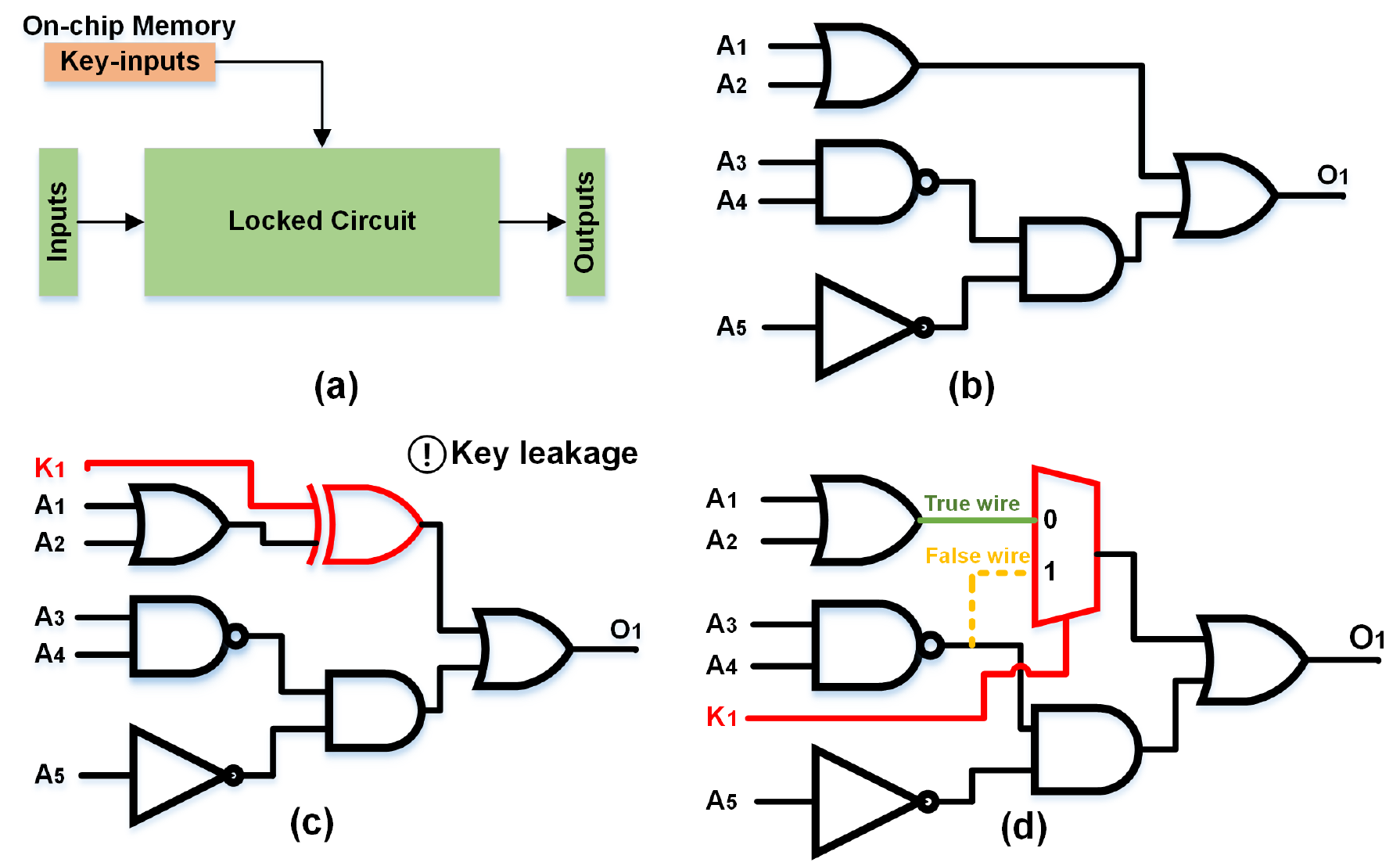}
  \caption{Logic locking techniques: (a) Overview; (b) Original netlist; (c)
XOR/XNOR-based logic locking; (d) MUX-based logic locking.}
  \label{fig:diff_logiclocking}
\end{figure}

Logic locking (LL) is a promising solution to protect a design's intellectual property (IP) from various threats throughout the integrated circuit (IC) supply chain~\cite{rostami2014primer}. LL performs functional and structural design alterations through the insertion of additional key-controlled logic such as XOR/XNOR gates, multiplexers (MUXes), or look-up tables (LUTs)~\cite{roy2010ending, plaza2015solving, chowdhury2021enhancing, hu2021risk}. Figure~\ref{fig:diff_logiclocking} shows examples from different LL methods. The  security of LL has, however, been challenged by the development of various attacks which can be broadly classified into two categories, namely, \emph{oracle-guided} (OG) and \emph{oracle-less} (OL). In OG attacks~\cite{subramanyan2015evaluating}, attackers access the locked design netlist and a functional chip, i.e., the \emph{oracle}. In OL attacks~\cite{chakraborty2018sail, sisejkovic2021challenging}, the attackers only have the locked design netlist at their disposal. OL attacks pose greater threats to LL as they can be mounted even if an oracle is not available.  

Among OL attacks, those based on machine learning (ML)~\cite{chakraborty2018sail,sisejkovic2021challenging} aim to predict the correct key by exploiting the information leakage occurring through the structural signatures induced by the different LL schemes. In fact, in XOR/XNOR-based and  AND/OR-based LL, there usually exists a direct mapping between the type of the key gate and the key value. Attacks like SAIL~\cite{chakraborty2018sail}, SnapShot~\cite{sisejkovic2021challenging}, and OMLA~\cite{alrahis2021omla} leverage this information along with the surrounding circuitry of a key gate to uncover the key value using deep learning models. MUX-based LL tends to leak less information  about the key value via the structure of the key gate. However, other attacks, like the constant propagation attacks SWEEP~\cite{alaql2019sweep} and SCOPE~\cite{alaql2021scope} and the structural analysis attack on MUX-based LL (SAAM), can successfully break MUX-based locking, calling for the development of learning-resilient LL techniques, such as deceptive MUX-based (D-MUX)~\cite{sisejkovic2021deceptive} LL and symmetric MUX-based LL~\cite{alaql2021scope}.

Unfortunately, however, a new attack based on graph neural networks (GNNs), called MuxLink~\cite{alrahis2022muxlink}, has been recently  developed that can break D-MUX and symmetric MUX-based locking and recover the correct key with 100\% accuracy for many designs from the ISCAS-85 and ITC-99 benchmarks. MuxLink maps the task of finding the correct key to a link-prediction problem, i.e., the problem of predicting the likelihood for a link (or wire) to be present in a design. In link prediction~\cite{zhang2018link}, a local subgraph is extracted around each (true or false) link, which is then processed by GNN layers to learn the mapping from subgraph patterns to the likelihood of existence of the link. Overall, a MUX-based locking technique that is resilient to the existing ML-based attacks remains elusive.

In this paper, we present SimLL, a novel similarity-based logic locking technique that locks a design using MUXes and shows robustness against the state-of-the-art OL ML-based attacks~\cite{alrahis2022muxlink, sisejkovic2021challenging} and constant propagation attacks~\cite{alaql2019sweep, alaql2021scope}. We observe that, in D-MUX or symmetric MUX-based locking, the local subgraph structure associated with the false link may be significantly different from that of the true link, which is exploited by MuxLink to train a GNN model that can differentiate between the true and false wires in a design. 
We may then mitigate this vulnerability by choosing the false inputs to the MUX in such a way that the local subgraph structure associated with them is topologically and functionally ``similar'' to that of the true wires. SimLL addresses this challenge by leveraging a notion of  topological and functional similarity of logic gates and wires to insert key-controlled MUXes. Topologically and functionally similar logic gates or wires are identified using a message passing-based algorithm and clustered into groups. Logic gates or wires from the same cluster are then used to insert key-controlled MUXes, resulting in true and false wires having similar subgraph structures. In summary, this paper makes the following contributions:
\begin{itemize}
    \item {A scalable, message passing-based graph theoretic approach to clustering topologically and functionally similar logic gates or wires in a design.}
    \item{SimLL, an LL scheme that inserts MUXes between topologically and functionally similar gates or wires and shows resilience against existing ML-based attacks.}
\end{itemize}
To the best of our knowledge, this is the first approach leveraging clusters of topologically and functionally similar logic gates and wires to robustify LL against ML-based attacks. We evaluate the effectiveness of SimLL against different ML-based OL attacks on various ISCAS-85 and ITC-99 benchmarks. Empirical results show that SimLL can reduce the average accuracy achieved by the ML-based attacks to  $\sim$50\%, bringing little to no advantage over a random guessing attack.

\section{Background and Related Work}\label{sec:background}

In this section, we discuss the existing MUX-based learning-resilient locking techniques and the OL attacks on them.

\begin{figure}[t]
  \centering
  \includegraphics[width=0.95\columnwidth]{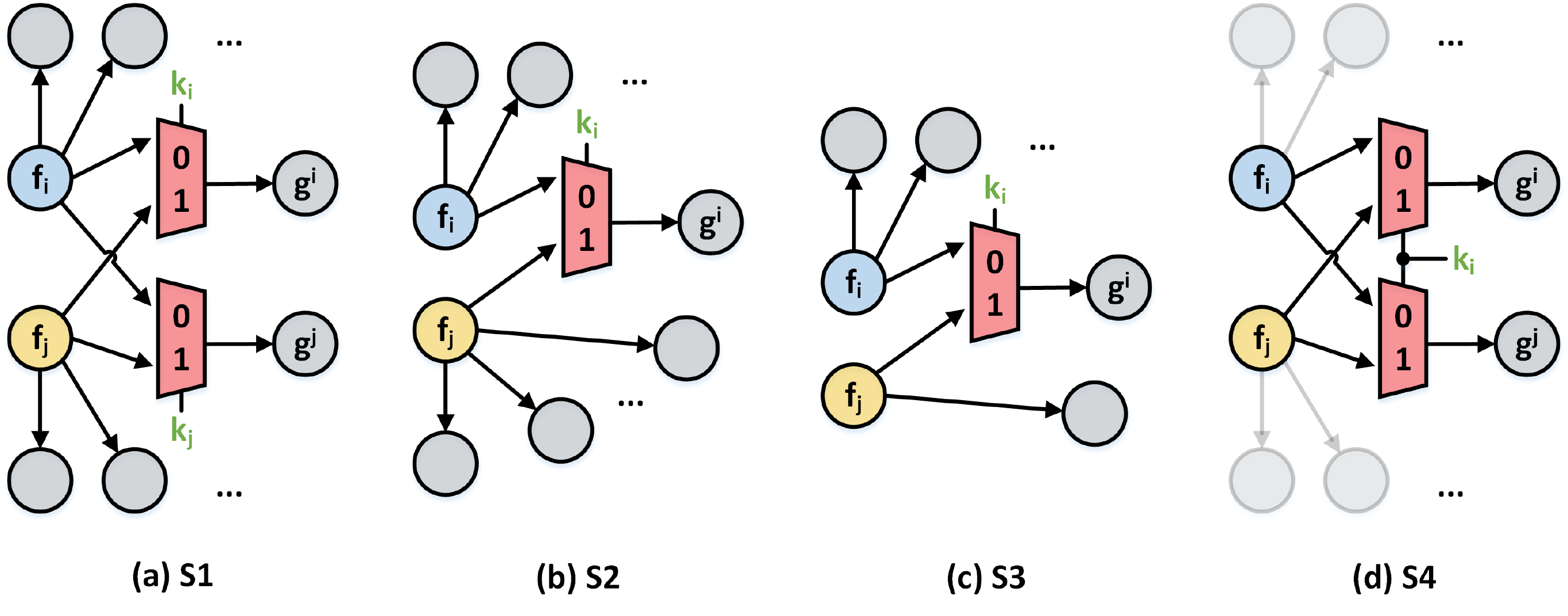}
  \caption{D-MUX locking strategies.}
  \label{fig:dmux}
  \vspace{-4mm}
\end{figure}

\subsection{Learning-Resilient MUX-Based Locking}

MUX-based locking inserts a MUX key-gate into a design, which takes in two wires from the circuit as inputs, while a key-input is connected to the select line, allowing the true wire to pass upon applying the correct key value. Differently from other locking schemes like XOR/XNOR or AND/OR insertion, MUX-based locking reconfigures the existing logic by always inserting the same element, i.e., a multiplexer. However, na\"{i}ve implementations of MUX-based locking are vulnerable to different OL attacks like SWEEP, SCOPE~\cite{alaql2019sweep,alaql2021scope}, and SAAM~\cite{sisejkovic2021deceptive}.

Recently, a learning-resilient MUX-based locking technique, D-MUX, has been proposed which successfully thwarts all the above attacks. D-MUX ensures that each wire feeding the MUX has the same probability of being true or false. To do so, D-MUX introduces different locking strategies ($S1$-$S4$), as shown in Fig.~\ref{fig:dmux}. 
Multi-output gates are those which have fan-out of $2$ or more while single-output gates have fan-out of $1$. Strategy $S1$ selects two multi-output gates $(f_i, f_j)$. An output gate corresponding to each gate,  i.e., $(g^i, g^j)$ is selected randomly from the respective fan-out gates. Next, two MUXes, controlled by key inputs $(k_i, k_j)$, are introduced between the two input gates $(f_i, f_j)$ and output gates $(g^i, g^j)$ to generate four valid paths for all values of $(k_i, k_j)$. Out of these four paths, $f_i \rightarrow g^i$ and $f_j \rightarrow g^j$ are the \emph{true paths} while  $f_i \rightarrow g^j$ and $f_j \rightarrow g^i$ are the \emph{false paths}. Since both input gates have multiple outputs, an attacker cannot determine which path is true or false. Strategy $S2$ selects two multi-output gates $(f_i, f_j)$ but performs locking using a single MUX controlled by a single key-input $k_i$. A randomly selected output gate for one of the input gates is locked, as shown in Fig.~\ref{fig:dmux}(b), where $S2$ selects $f_i$ and its output gate $g^i$.

Strategy $S3$ selects one multi-output gate $f_i$, which is then locked using a single key-input $k_i$ controlling one MUX. Given a single-output gate $f_j$, an output gate is selected from the multi-output input gate's fan-out so that, regardless of the key bit value, input gates $(f_i, f_j)$ remain connected to the circuit. Finally, strategy $S4$ selects two gates $(f_i, f_j)$, which can be multi- or single-output gates, and introduces two MUXes which are controlled by a single key-input $k_i$. In all these strategies, the MUXes are configured in such a way that, regardless of the key bit value, both the input gates always remain connected to the circuit, i.e., there are no floating gates, and no combinational loops are created. Symmetric MUX-based locking~\cite{alaql2021scope} adopts a strategy similar to $S4$ of D-MUX, but instead of a single key input, two key inputs control the two MUXes.

The SAAM~\cite{sisejkovic2021deceptive} attack analyzes the logic gates connected to a MUX-key gate to identify the true wire. The attacker iterates over each key bit and applies the values $0$ and $1$ to check if any of these results in a netlist with structural faults. For example, for the circuit in Fig.~\ref{fig:diff_logiclocking}(d), the attacker applies $0$ to the key input, which forwards the true wire to the output, and the resulting netlist displays no structural faults. However, applying a $1$ forwards the false wire to the output and the true wire remains floating. In a functional design, we do not have any floating internal wires. The attacker can then identify the MUX input that remains floating as the true wire. D-MUX is resilient to SAAM since every strategy in D-MUX makes sure that, regardless of the key bit value, both the input gates of the MUX remain connected to the circuit via valid paths~\cite{sisejkovic2021deceptive}.

A constant propagation attack like SWEEP~\cite{alaql2019sweep,alaql2021scope} hard-codes the value of a single key input at a time and performs re-synthesis to extract relevant design features (like area and power consumption) from the two synthesis reports corresponding to each key value. Next, it learns a set of weight variables that captures the correlation between the correct key value and the features from the training dataset. During test, SWEEP uses the learned weights and the target features to deduce a value for each key input which can be $0$, $1$,
or $X$, where $X$ means that the attack was not able to make a decision.  
SWEEP has an adjustable margin $m$ which can be tuned by an attacker. The default value of $m$ is $0$. In the case of D-MUX, the features resulting from the two key input values are similar, since no circuit reduction occurs for wrong key-input values, making it resilient to constant propagation attacks~\cite{sisejkovic2021deceptive}. 

\begin{figure}[t]
  \includegraphics[width=\columnwidth]{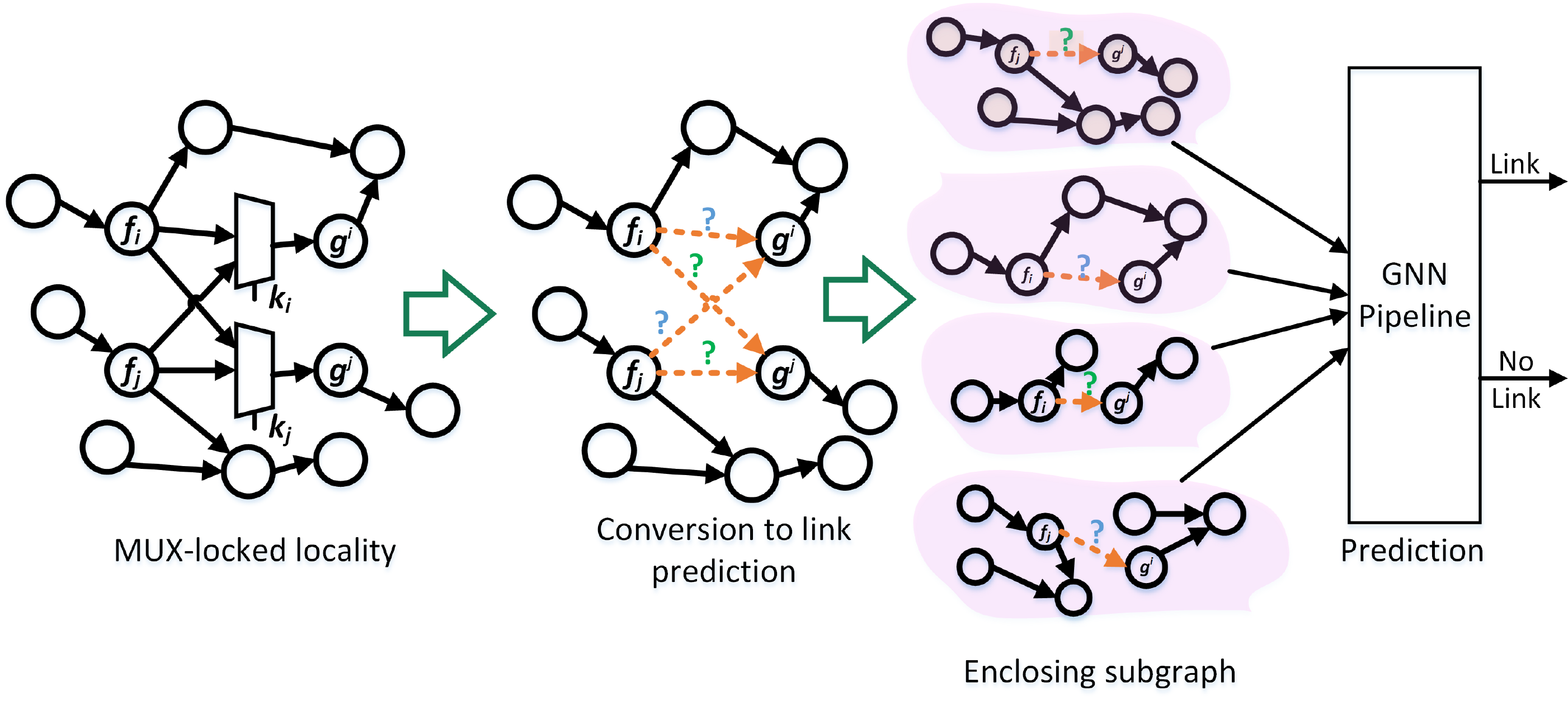}
  \caption{Workflow of the MuxLink attack. A MUX-locked locality, i.e., a fixed-depth subcircuit surrounding the MUX key gates, is  converted to a graph to solve the corresponding link prediction problem.}
  \label{fig:muxlink}
\end{figure}

\subsection{State-of-the-Art ML-Based Attack on D-MUX Locking}

The security guarantees of D-MUX locking have been challenged by MuxLink~\cite{alrahis2022muxlink}, an ML-based attack that can recover the correct key with approximately 100\% accuracy for different designs. MuxLink maps the task of finding the true MUX input into a link prediction problem. It trains a GNN model to predict the likelihood of a wire (link) to be present in a design by leveraging the structure of the given circuit and the interconnection of gates surrounding the true wires. Each MUX key gate in a locked design is associated with a true and false link. During inference, for each of these links, MuxLink, extracts a subgraph and processes it with the trained GNN model to predict the likelihood score for the link to be present in the design. If the likelihood score of the true wire is significantly higher than the one of the false wire, MuxLink can indeed identify the true wire. 
If the difference between the likelihood scores of the wires is less than a certain threshold $(\emph{th})$, then MuxLink predicts the key bit value associated with the MUX as $X$, implying that it is not able to make a decision. Fig.~\ref{fig:muxlink} shows the workflow for MuxLink. 

\section{SimLL: Similarity-Based Logic Locking}\label{sec:simll}

We first specify our threat model. Then, we discuss how structural and functional properties of logic gates and wires in a design can be used to introduce false wires that confuse the MuxLink GNN model. 
Finally, we introduce SimLL. 

\subsubsection*{\textbf{Threat Model}}

{We assume that the adversary: $(a)$ has access to the locked netlist only (OL model); $(b)$ knows the algorithmic details of the LL scheme; $(c)$ knows the location of the key inputs in the netlist.}

\subsubsection*{\textbf{Preliminaries}}

D-MUX LL considers the fan-out of the gates when choosing the inputs to a MUX key gate. However, the location of the input gates to the MUX are chosen randomly while making sure that no combinational feedback loop is introduced. This may result in the true and the false wires having significantly different subgraphs associated with them, making it easy for the GNN model to differentiate between them. We can mitigate this weakness by choosing the inputs to the MUX in such a way that the induced subgraphs are ``similar'' to make it difficult for the GNN model to discriminate between true and false wires. We address this challenge by searching for topological and functional similarities in wires and gates. Among the few approaches in the literature for detecting structural and functional similarities in a design, some efforts focus on finding similar registers~\cite{geist2020relic, chowdhury2021reignn}, but their extension to generic 
gates and wires is not straightforward. We propose instead two algorithms inspired by a classic result for testing graph isomorphism~\cite{shervashidze2011weisfeiler}.

\subsection{Clustering Topologically and Functionally Similar Nodes}

\setlength{\textfloatsep}{0pt}
\begin{algorithm}[t]
    \scriptsize
	\caption{Node Clustering} 
	\label{alg:cls_nodes} 
		\textbf{Input: } {$G =(V,E,X_V)$: graph generated from the netlist, $h=$ \# hops} \\
    \textbf{Output: }{$NC$: node clusters} 
	\begin{algorithmic}[1]
	\State $s_{v}^{0} \gets \Call{Update}{X_v}, \forall v \in V$
	\For {$k \gets 1$ to $h$}
   \State $s_{nv}^{k-1} \gets \Call{Sort}{\{\{s_{w}^{k-1}: w \in N_{G}(v) \}\}}, \forall v \in V$
   \State $s_{v,nv}^{k-1} \gets \Call{Concat}{s_{v}^{k-1}, s_{nv}^{k-1}}, \forall v \in V$
    \State $s_{v}^k \gets \Call{Update}{s_{v,nv}^{k-1}}, \forall v \in V$
	\EndFor
        \State \Return \Call{node\_cluster}{\{\{$s_v^h, \forall v \in V$\}\}}
        \State
        \Function{node\_cluster}{$S$}
            \State $NC \gets \{\}$
            \For {$s_v \in S$}
                \If {$s_v = s_n$, where $n$ is from $c \in NC$}
                    \State add $v$ into $c$
                \Else
                    \State create a new set $c'=\{v\}$ and add it into $NC$
                \EndIf
            \EndFor  
            \State \Return $NC$
        \EndFunction
	\end{algorithmic}
\end{algorithm}

We convert the circuit netlist to a graph $G = (E, V, X_V)$, where $V = \{1, \ldots, n\}$
is the set of nodes (gates), $E \subseteq V \times V$ is the set of links (wires), and $X_V$ is the set of node features.  Each node $v$ is associated with a feature $X_v$ consisting of the logic gate type encoded as a string. One-hop neighbors of node $v$ are denoted by $N_{G}(v) = \{w : (v, w) \in E\}$. We use  $\{\{\cdot\}\}$ to denote a multiset.

As outlined in Algorithm~\ref{alg:cls_nodes}, each node is also associated with a state $s_v$ that encodes as a string the structural and functional properties of the node and its neighborhood. The state of a node is computed using the \Call{Update}{} function, an injective  hash function that takes a string as an input and returns another string as the output. \Call{Update}{} has the following property: for any pair of strings $v, w$, $\Call{Update}{v} =  \Call{Update}{w}$ if and only if $ v = w$. The initial state of a node is calculated as $s_{v}^{0} = $\Call{Update}{$X_v$} and  
subsequently updated iteratively by aggregating the states of the neighboring nodes $N_{G}(v)$.  
For each node $v$, its current state is concatenated with the current states of the nodes in $N_{G}(v)$ and used to generate the new state using the \Call{Update}{} function. 
Depending on the number of hops $h$, each node aggregates information from neighbors which are at a distance less than or equal to $h$ from it. Finally, nodes that end up having the same state are marked as \emph{similar} and clustered together, meaning that they possess the same $h$-hop neighborhood. 

\subsection{Clustering Topologically and Functionally Similar Links}

The procedure for clustering similar links is outlined in Algorithm~\ref{alg:cls_edges}. We first extract an $h$-hop enclosing subgraph $g_{(u,v)}$ for each link $(u, v)$. We call the link under consideration the target link, and the nodes connected by it the target nodes. The subgraph $g_{(u,v)}$ is obtained from $G$ and contains the nodes $ \{ x | d(u, x) \leq h$ or $d(v, x) \leq h\}$, where $d(u,x)$ is the distance between nodes $u$ and $x$. 
To distinguish between target nodes and other nodes in the subgraph, a new feature $l$ is added to each node in addition to the gate type. The string $l$ is assigned using the double radius node labeling (DRNL)~\cite{zhang2018link} technique. The DRNL label for node $x$ is calculated as $l_x = 1 + \min (d_u, d_v) + (d/2)[(d/2) + (d\%2) -1]$ where $d_u = d(x,u)$, $d_v = d(x,v)$ and $d = d_u + d_v$. $d/2$ and $d\%2$ are the integer quotient and remainder of $d$ divided by 2, respectively. If $x$ has a path to only one of the target nodes, then $l_x = 0$. The target nodes have $l_x = 1$. DRNL assigns larger labels to nodes with larger distances with respect to both target nodes. 

Similarly to Algorithm~\ref{alg:cls_nodes}, each node $x$ in the subgraph has a state $s_x$ which is then updated by aggregating the states of the neighboring nodes. At the end of $h$ hops, the subgraph corresponding to each link is defined by a multiset  of node states, denoted by $H_{(u,v)}$. We then cluster links that have the same multisets together, where two multisets are called equal if they have the same elements with equal multiplicities~\cite{shervashidze2011weisfeiler}. Having the same multisets indicates that the links possess the same $h$-hop neighborhood.   

\setlength{\textfloatsep}{0pt}
\begin{algorithm}[t]
    \scriptsize
	\caption{Link Clustering} 
	\label{alg:cls_edges} 
		\textbf{Input: } {$G =(V,E,X_V)$: graph generated from the netlist, $h =$ \# hops} \\
    \textbf{Output: }{$LC$: link clusters} 
	\begin{algorithmic}[1]
	\State $\text{Extract $h$-enclosing subgraph } g_{(u,v)} \forall \ (u,v) \in E$
	\For {each $g_{(u,v)}$}
	\State $X_a' \gets \Call{Concat}{X_a, l_a}, \forall \ a \in g_{(u,v)}$
	\State $s_{a}^{0} \gets \Call{Update}{X_a'}, \forall \ a \in g_{(u,v)}$
	\For {$k \gets 1$ to $h$}
          \State $s_{na}^{k-1} \gets \Call{Sort}{\{\{s_{w}^{k-1}: w \in N_{g_{(u,v)}}(a)\}\}}, \forall a \in g_{(u,v)}$
          \State $s_{a, na}^{k-1} \gets \Call{Concat}{s_{a}^{k-1}, s_{na}^{k-1}}, \forall a \in g_{(u,v)}$
	   \State $s_{a}^k \gets \Call{Update}{s_{a, na}^{k-1}}, \forall a \in g_{(u,v)} $

	\EndFor
	\State $H_{(u,v)} = \{\{s_{a}^h: \forall a \in g_{(u,v)} \}\}$
        \State \Return \Call{link\_cluster}{\{\{$H_{(u,v)}$\}\}}
	\EndFor
        \State
        \Function{link\_cluster}{$H_G$}
            \State $LC \gets \{\}$
            \For {$h_{g(e)} \in H_G$}
                \If {$h_{g(e)}$ = $h_{g(n)}$, where $n$ is from $c \in LC$}
                    \State add $e$ into $c$
                \Else
                    \State create a new set $c'=\{e\}$ and add it into $LC$
                \EndIf
            \EndFor  
            \State \Return $LC$
        \EndFunction
	\end{algorithmic}
\end{algorithm}

\begin{algorithm}[t]
    \scriptsize
	\caption{SimLL Scheme} 
	\label{alg:simll} 
		\textbf{Input: } {Netlist $N$, key $K$, available locking strategies $L_s$, hop \# $h$} \\
    \textbf{Output: }{Locked netlist} 
	\begin{algorithmic}[1]
	\State $\text{Convert } N \text{ to graph } G =(V,E,X_V) $
	\State $ node\_clusters \gets \Call{Node\_Clustering}{G(V,E,X_V), h}$
	\State $ link\_clusters \gets \Call{Link\_Clustering}{G(V,E,X_V), h}$
        \State $kl \gets \Call{Len}{K}$ \Comment{The number of unused key inputs}
	\For {$c \in link\_clusters$} \Comment{Introduce MUXes for similar links}
            \State divide $c$ into $E_{s}^{c}$ and $E_{m}^{c}$
            \While {$kl \geq 2$ and $|E_{m}^{c}| \geq 2$}
                \State select two links $l_i, l_j \in E_{m}^{c}$
                \State apply $S1$ on $N$ with $l_i, l_j$ and then remove $l_i, l_j$ from $E_{m}^{c}$
            \EndWhile
            \While {$kl \geq 1$ and $|E_{s}^{c}| \geq 2$}
                \State select two links $l_i, l_j \in E_{s}^{c}$
                \State apply $S4$ on $N$ with $l_i, l_j$ and then remove $l_i, l_j$ from $E_{s}^{c}$ 
            \EndWhile
        \EndFor
	\For {$c \in node\_clusters$} \Comment{Introduce MUXes for similar nodes}
            \State divide $c$ into $F_{s}^{c}$ and $F_{m}^{c}$
            \While {$kl \geq 1$ and $\left(|F_{m}^{c}| + |F_{s}^{c}| \geq 2\right)$}
                \State \Call{RandomShuffle}{$L_s$} 
                \For {$S_i \in L_s$}
                    \If {$S_i = S1$ and $|F_m^c| \geq 2$ and $kl \geq 2$}
                        \State select two nodes $f_i, f_j \in F_m^c$ and then break
                    \ElsIf {$S_i = S2$ and $|F_m^c| \geq 2$}
                        \State select two nodes $f_i, f_j \in F_m^c$ and then break
                    \ElsIf {$S_i = S3$ and $|F_m^c| \geq 1$ and $|F_s^c| \geq 1$}
                        \State select two nodes $f_i \in F_m^c, f_j \in F_s^c$ and then break
                    \ElsIf {$S_i = S4$}
                        \State select two nodes $f_i, f_j \in \left(F_m^s \cup F_s^c\right)$ and then break
                    \EndIf
                \EndFor
                \State randomly select  $g^i, g^j$ s.t. $(f_i, g^i) \in E, (f_j, g^j) \in E$
                \State apply $S_i$ on $N$ with $f_i, f_j, g^i, g^j$
            \EndWhile
        \EndFor
        \While {$kl \geq 1$} \Comment{Run D-MUX if unused key inputs exist}
            \State $\Call{D-MUX}{L_s, K, N}$
        \EndWhile
        \State \Return $N$
	\end{algorithmic}
\end{algorithm}

\begin{figure*}[htbp]
  \centering
  \includegraphics[width=0.95\linewidth]{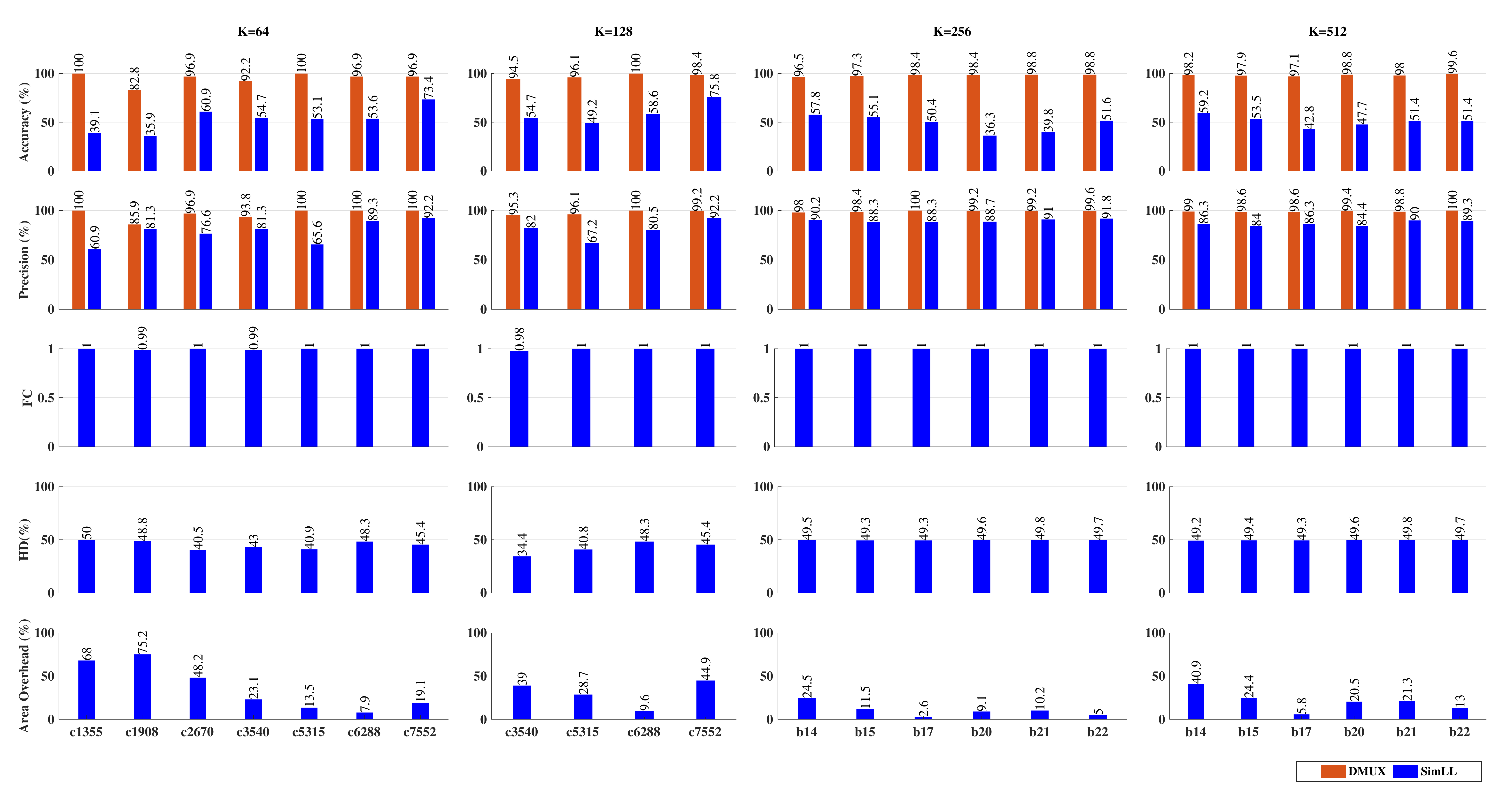}
  \vspace{-15pt}
  \caption{MuxLink attack evaluation on SimLL locked benchmarks. ISCAS-85 benchmarks are locked with 64 and 128 key bits while ITC-99 benchmarks are locked with 256 and 512 keys. Success of the attack is evaluated using AC, PC, FC, and HD. AC and PC achieved by MuxLink on D-MUX locked designs are also included for comparison. Area overhead of SimLL is also reported.}
  \label{fig:simll_attack_muxlink_ac_pc_fc_hd_area}
\end{figure*}

\subsection{SimLL Scheme}

In Algorithm~\ref{alg:simll} SimLL leverages the different locking strategies $S1 - S4$ of D-MUX to introduce the MUX key gates. However, instead of choosing the inputs to the MUX gates randomly, it follows a systematic approach. Algorithm~\ref{alg:simll} outlines  SimLL, which takes as inputs (a) the original netlist $N$, (b) the correct key vector $K$, (c) a set of available strategies $L_s$, and (d) number of hops $h$ for subgraph extraction.

SimLL calls Algorithm~\ref{alg:cls_nodes} and~\ref{alg:cls_edges} to get (i) clusters of similar nodes and (ii) clusters of similar links.
It then introduces MUX key gates by first leveraging similar links, followed by similar nodes. Once the clusters of similar links are obtained, the links in a single cluster are further divided into two sets, namely single-output source links $E_{s}^c$ and multi-output source links $E_{m}^c$, where $E_{s}^c$ contains all those links in cluster $c$ whose source node has fan-out of 1 while $E_{m}^c$ contains links whose source node has fan-out of 2 or more. SimLL starts performing locking by choosing two links $l_i, l_j$ from the $E_{m}^c$ set and introduces two MUXes $M_1, M_2$ controlled by two key inputs $k_i, k_j$ following strategy $S1$ while making sure that no combinational loops are introduced. $M_1$ will have nodes  $f_i$ and $ f_j$ connected to its inputs, $k_i$ to its control input, and $g^i$ connected to its output. For $M_2$, $f_i$ and $f_j$ are connected to its inputs, $k_j$ to the control input, and $g^j$ to its output. SimLL iterates over $E_{m}^c$ and introduces the MUXes until the key inputs are exhausted. If there still exist more key inputs, then SimLL performs locking by taking links from  $E_{s}^c$ and introducing two MUXes controlled by a single key input following strategy $S4$. $S4$ is similar to $S1$ except that only one key input $k_i$ controls both the MUXes and the source nodes can be single or multi-output.

In case all the edge clusters are exhausted while key inputs are still left, SimLL leverages similar nodes in the design. Before performing locking, SimLL divides nodes in a single cluster into a single-output node set $F_{s}^c$ and a multi-output node set $F_{m}^c$. Then, it iterates over the clusters and randomly shuffles the available strategies to pick a candidate strategy and suitable nodes $f_i, f_j$ from the cluster. SimLL needs to choose valid output nodes $g^i, g^j$ for $f_i, f_j$ from their fan-out nodes. If the chosen output nodes do not introduce any combinational feedback loop in the design, SimLL goes ahead and introduces single MUXes between the input nodes $f_i,f_j$ and output node $g^i$, in the of case strategies $S2$ or $S3$, or two MUXes in the case of $S1$ and $S4$. If all the node clusters are used up and some unused key inputs still exist, SimLL introduces the MUX randomly following the D-MUX locking approach.

\section{Evaluation}\label{sec:evaluation}

We evaluate SimLL against MuxLink~\cite{alrahis2022muxlink}, SWEEP~\cite{alaql2019sweep}, and SnapShot~\cite{sisejkovic2021challenging} for different designs from the ISCAS-85 and ITC-99 benchmarks. The ISCAS-85 benchmarks are locked with 64 and 128 keys while the ITC-99 benchmarks are locked with 256 and 512 keys to assess the impact of different key sizes on the performance of the existing attacks on SimLL. We used the 45-nm Nangate Open Cell Library to synthesize the designs.
In SimLL, we set the value of $h$ to $2$. In fact, higher values of $h$ tend to result in smaller cluster sizes, which leads to fewer options for introducing MUXes between similar nodes or links, making SimLL less effective.
We implemented SimLL in Python and performed all the experiments on a server with an NVIDIA GeForce RTX 2080 graphics card, 48 2.1-GHz processor cores, and 500-GB memory.

\subsubsection{Evaluation Metrics}

We use the following metrics for evaluating the resilience of SimLL against different types of attacks: key accuracy (AC), precision (PC), functional corruptibility (FC), and Hamming distance (HD). AC is defined as the percentage ratio between the correctly predicted key bits $K_{correct}$ and the total number of key bits $K_{total}$, i.e.,  $\text{AC} = (K_{correct}/K_{total}) \times 100$. PC measures the ratio of the correctly predicted key bits, counting every $X$ as a correct guess, i.e., $PC = ((K_{correct} + K_{X})/K_{total}) \times 100$, where $K_{X}$ is the number of key bits for which the attack is unable to make a decision. FC represents the ratio between the number of input patterns for which the output is incorrect and the number of applied input patterns along with the predicted key. HD measures the average number of incorrect bits in the output. They are defined as follows, 
\begin{equation}
    \resizebox{0.5\hsize}{!}{$\text{FC}=\frac{1}{|I|}\sum_{i \in I} \mathbbm{1}(O_{X,i} \neq O_{C,i})$},
\end{equation}
\begin{equation}
    \resizebox{0.56\hsize}{!}{$\text{HD}=\frac{1}{|I|}\sum_{i \in I} \sum_{j \in m} O_{X,i,j} \oplus O_{C,i,j} $},
\end{equation}
where $O_{X,i}$ is the output given the predicted key and an input pattern $i$, $O_{C,i}$ is the output given the correct key and an input pattern $i$, $O_{X,i,j}$ is the $j$-th bit of $O_{X,i}$, and $O_{C,i,j}$ is the $j$-th bit of $O_{C,i}$. To measure the FC and HD, we simulate the locked and oracle designs with $|I| = 200,000$ random input patterns.

\subsubsection{MuxLink}

We evaluate the resilience of SimLL against MuxLink using the open-source MuxLink tool~\cite{alrahis2022muxlink}. We use the same GNN model architecture defined in MuxLink~\cite{alrahis2022muxlink}, consisting of four graph convolution (GCN) layers, two $1-D$ convolutional layers, a single fully connected layer, a dropout layer, and a softmax layer of 2 output units. A $\tanh$ activation function is used for the GCN layers while the $\mathrm{ReLU}$ function is used in the rest of the layers. Similar to MuxLink, we set the threshold $\emph{th}$ to be $0.01$. For each benchmark design, we lock it using SimLL and then perform the MuxLink attack. The AC, PC, FC, and HD achieved by MuxLink on SimLL locked benchmarks are presented in Fig.~\ref{fig:simll_attack_muxlink_ac_pc_fc_hd_area}. We also include the performance of MuxLink on D-MUX locked circuits for comparison. 

MuxLink achieves average AC of $97.1\%$ and PC of $98.1\%$ on benchmarks locked with D-MUX~\cite{alrahis2022muxlink}. However, when we apply the attack on designs locked with SimLL, the AC and PC reduce to $52.4\%$ and $83.8\%$, respectively. In the case of D-MUX, the difference between PC and AC is $1\%$, implying that the number of key bits for which the attack is unable to make a decision is very low. The difference between PC and AC increases to $31.4\%$ for SimLL, meaning that a large number of key bits are classified as $X$. SimLL is then successful in confusing the ML model. Moreover, the average PC of $83.9\%$ shows that a large number of key inputs are incorrectly predicted by MuxLink and SimLL is successful in making the ML model incorrectly identify some false links as true links.  While the AC reduces by $\sim$40\% for most of the designs, $c7552$ provides $\sim$25\%, mostly due to the fact that the number of similar nodes and links in the clusters was smaller than in other benchmarks and many MUX inputs were chosen randomly as in D-MUX.  

{We also measure the FC and HD for the key predicted by MuxLink on the SimLL locked circuits. Since MuxLink predicts many key bits to be $X$, to mimic the decision of an attacker, we set the value for these key bits randomly, averaged over multiple runs. The average FC is $0.99$ and the average HD is $46.94\%$ across the different SimLL locked designs, showing that the design recovered by the MuxLink attack can exhibit a significant amount of errors with respect to the original design. Figure~\ref{fig:simll_attack_muxlink_ac_pc_fc_hd_area} shows the area overhead incurred by SimLL for the different benchmarks, which is similar to the one of D-MUX.}

\begin{figure}[t]
  \includegraphics[width=1.05\columnwidth]{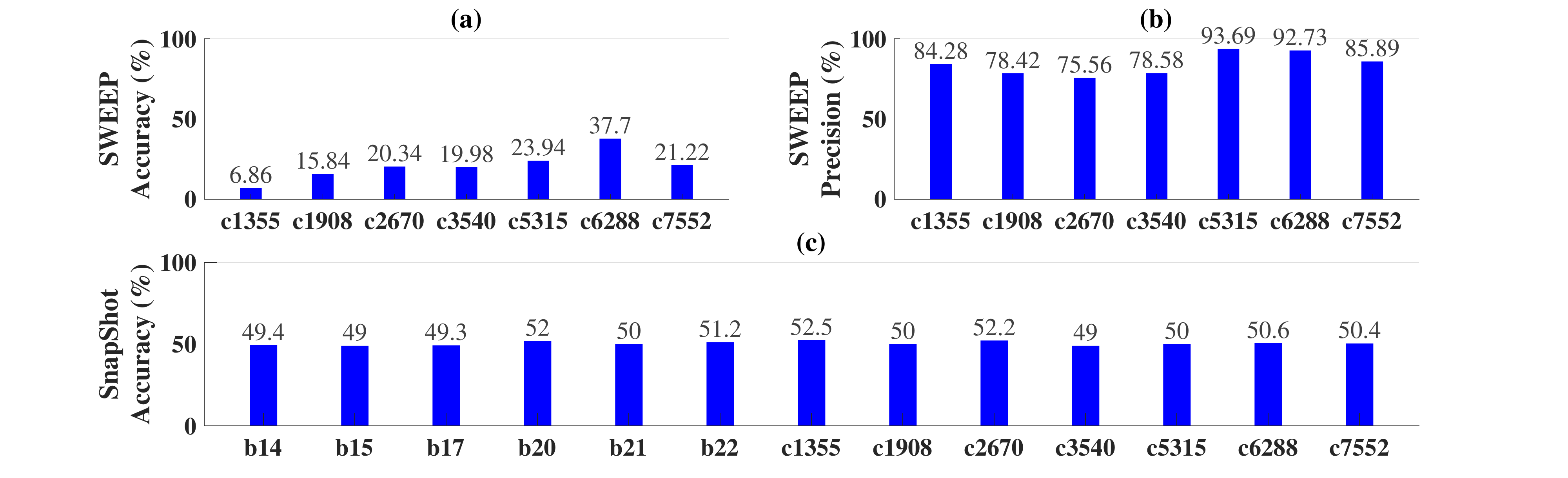}
  \caption{(a)(b) SWEEP attack evaluation on SimLL. (c) SnapShot attack evaluation on SimLL under the SRS attack scenario.}
  \label{fig:sweep_snapshot}
\end{figure}

\subsubsection{SWEEP}

We evaluate the resiliency of SimLL to the constant propagation attack SWEEP~\cite{alaql2019sweep} using the open-source tool. We follow the same setup used in the literature~\cite{sisejkovic2021deceptive}. Seven ISCAS-85 benchmarks were copied 100 times and locked with SimLL with a key size of 64, resulting in a data set of 700 locked benchmarks. For each benchmark, the data set is divided into a test set consisting of the 100 locked designs of the benchmark in question and a training set consisting of the remaining 600 other designs. 
We recall that SWEEP predicts the value of a key bit as $0$, $1$, or $X$. We set the value of the margin $m$ to be initially $0$, which results in an average AC of $21\%$ and an average PC of $84.2\%$, as shown in Fig.~\ref{fig:sweep_snapshot} (a) and (b). SimLL is then resilient to SWEEP. When the margin $m$ is slightly increased, AC falls to $0\%$ while PC rises to $100\%$, which implies that SWEEP predicts all the key bits as $X$.

\subsubsection{SnapShot}

We also evaluate the resiliency of SimLL to the ML-based OL attack SnapShot using the self-referencing scheme (SRS) attack scenario. SnapShot predicts the key bit value either as $0, 1$ by looking into locality vectors that represent the neighboring sub-circuit associated with the key gate. For each SimLL locked benchmark, a training
set is created by relocking the target netlist 100 times which is used for training a convolution neural network (CNN) model~\cite{sisejkovic2021challenging}. During inference, the trained model is used for predicting the key of the target locked netlist. As shown in Fig.~\ref{fig:sweep_snapshot} (c), SnapShot achieves an average AC of $\sim$50\%.

\section{Conclusion}\label{sec:conclusion}

We presented SimLL, a locking technique that introduces key-controlled MUXes between topologically and functionally similar logic gates and wires in a design to confuse ML-based attacks. Empirical results show that SimLL is resilient against the state-of-the-art OL attacks. Future work includes investigating improvements to SimLL when the available number of similar gates and wires in a design is small.

\section*{Acknowledgments}
This work was supported by the Air Force Research Laboratory (AFRL) and the Defense Advanced Research Projects Agency (DARPA) under agreement number FA8650-18-1-7817.

\bibliographystyle{ieeetr} 
\bibliography{references}

\end{document}